\documentclass[preprint,aps]{revtex4}
\usepackage{mathrsfs}

\usepackage{graphicx}
\usepackage{multirow}
\usepackage{makecell}
\usepackage{booktabs}
\usepackage{hyperref}
\begin{document}

\title{Absence of BCS-BEC Crossover in FeSe$_{0.45}$Te$_{0.55}$ Superconductor}

\author{Junjie Jia$^{1,2}$, Yadong Gu$^{1,2}$, Chaohui Yin$^{1,2}$, Yingjie Shu$^{1,2}$, Yiwen Chen$^{1,2}$, Jumin Shi$^{1,2}$, Xing Zhang$^{1,2}$, Hao Chen$^{1,2}$, Taimin Miao$^{1,2}$, Xiaolin Ren$^{1,2}$, Bo Liang$^{1,2}$, Wenpei Zhu$^{1,2}$, Neng Cai$^{1,2}$, Fengfeng Zhang$^{4}$, Shenjin Zhang$^{4}$, Feng Yang$^{4}$, Zhimin Wang$^{4}$, Qinjun Peng$^{4}$, Zuyan Xu$^{4}$, Hanqing Mao$^{1,2,3}$, Guodong Liu$^{1,2,3}$, Zhian Ren$^{1,2,3}$, Lin Zhao$^{1,2,3*}$ and X. J. Zhou$^{1,2,3,*}$}

\affiliation{
\\$^{1}$National Lab for Superconductivity, Beijing National Laboratory for Condensed Matter Physics, Institute of Physics,
Chinese Academy of Sciences, Beijing 100190, China
\\$^{2}$University of Chinese Academy of Sciences, Beijing 100049, China
\\$^{3}$ Songshan Lake Materials Laboratory, Dongguan 523808, China
\\$^{4}$Technical Institute of Physics and Chemistry, Chinese Academy of Sciences, Beijing 100190, China
\\$^{*}$Corresponding author: lzhao@iphy.ac.cn and XJZhou@iphy.ac.cn
}

\date{\today}

\maketitle

\noindent {\bf{Abstract}}\\

{\bf
In iron-based superconductor Fe(Se,Te), a flat band-like feature near the Fermi level was observed around the Brillouin zone center in the superconducting state. It is under debate whether this is the evidence on the presence of the BCS-BEC crossover in the superconductor. High-resolution laser-based angle-resolved photoemission measurements are carried out on high quality single crystals of FeSe$_{0.45}$Te$_{0.55}$ superconductor to address the issue. By employing different polarization geometries, we have resolved and isolated the d$_{yz}$ band and the topological surface band, making it possible to study their superconducting behaviors separately. The d$_{yz}$ band alone does not form a flat band-like feature in the superconducting state and the measured dispersion can be well described by the BCS picture. We find that the flat band-like feature is formed from the combination of the d$_{yz}$ band and the topological surface state band in the superconducting state. These results reveal the origin of the flat band-like feature and rule out the presence of BCS-BEC crossover in Fe(Se,Te) superconductor.
}\\


\newpage

\vspace{3mm}

\noindent {\bf{1. Introduction}}\\

The Bardeen-Cooper-Schrieffer (BCS) theory of superconductivity has been successful in explaining conventional superconductors in the weak coupling limit where electrons pair up to form bosons in the momentum space\cite{JRSchrieffer1957JBardeen}. For the atom gas system in the strong coupling limit, the Bose-Einstein-condensation (BEC) can occur at ultra-low temperature where the cold atoms condensate to a single quantum state\cite{SStringari1999FDalfovo,SStringari2008SGiorgini}.
In an intermediate regime between BCS and BEC, a BCS-BEC crossover is present which is characterized in a significantly different way from the strict BCS theory\cite{eagles1969possible,leggett2008diatomic,nozieres1985bose,KLevin2005QJChen,randeria2014crossover,chen2024superconductivity}. Recently, signatures of the BCS-BEC crossover are reported in some iron based superconductors where the Fermi energy $\varepsilon_F$ is small and comparable to the superconducting gap $\Delta$\cite{lubashevsky2012shallow,kasahara2014field,okazaki2014superconductivity,rinott2017tuning,hashimoto2020bose}.
In particular, there have been some reports from angle-resolved photoemission (ARPES) measurements to observe spectroscopic evidence of BCS-BEC crossover in FeSe$_x$Te$_{1-x}$\cite{lubashevsky2012shallow,okazaki2014superconductivity,rinott2017tuning}. In the superconducting state, a flat band-like feature is formed near the Fermi level around the Brillouin zone center which was taken as a signature of the BCS-BEC crossover in FeSe$_x$Te$_{1-x}$\cite{lubashevsky2012shallow,rinott2017tuning}. 
Such observations and the origin of the flat band-like feature have been controversial\cite{lubashevsky2012shallow,okazaki2014superconductivity,rinott2017tuning} and further ARPES investigation are needed to clarify these issues.

In this paper, we carried out high-resolution laser ARPES measurements on high quality FeSe$_{0.45}$Te$_{0.55}$  to resolve the detailed electronic structures in the normal and superconducting states. Different polarization geometries are used to disentangle the band structures around the zone center by taking advantage of the matrix element effects.
In the normal state, there are two bands crossing the Fermi level around zone center: the bulk d$_{yz}$ band and a nontrival topological surface state. In the superconducting state, the measured dispersion of the d$_{yz}$ band can be well described by the BCS picture. We find that the flat band-like feature is formed from the combination of the d$_{yz}$ band and the topological surface state band in the superconducting state. These results rule out the presence of BCS-BEC crossover in Fe(Se,Te) superconductor.\\

\noindent {\bf{2. Experiment}}\\

The Mn-substituted Fe$_{1-x}$Mn$_x$Se$_{0.45}$Te$_{0.55}$  crystals were grown by using self-flux method\cite{ZARen2022YDGu}. The replacement of Fe with minor Mn can effectively suppress the formation of the interstitial iron atoms and significantly improve the single crystal quality and superconducting properties (Fig. 1a)\cite{ZARen2022YDGu}.
The samples used in this work are Fe$_{0.98}$Mn$_{0.02}$Se$_{0.45}$Te$_{0.55}$ (named as FeS45T55) which show a superconducting  transition temperature T$_C$ of $\sim$14.2\,K, with a transition width of $\sim$1.5\,K from the magnetic susceptibility measurement (inset in Fig. 1a).
High resolution angle-resolved photoemission measurements were carried out on our ARToF-ARPES system which is equipped with the  vacuum-ultra-violet (VUV) laser (h$\nu$=6.994\,eV) and angle-resolved time-of-flight (ARToF) electron energy analyzer\cite{XJZhou2018}. The ARToF-ARPES can cover two-dimensional momentum space simultaneously and has much weaker non-linearity effect so that the intrinsic signal can be measured. The energy resolution was set at 1\,meV and the angular resolution was 0.3$^\circ$ corresponding to 0.004${\AA}^{-1}$ momentum resolution at the photon energy of 6.994 eV. The sample was cleaved at 6\,K and measured in ultrahigh vacuum with a base pressure about 5$\times$10$^{-11}$ mbar. The Fermi level is referenced by measuring on a clean polycrystalline gold that is electrically connected to the sample, as well as the normal state ARPES measurements of the sample.\\

\noindent {\bf{3. Results and discussions}}\\

Figure 1 shows the overall Fermi surface mappings of the FeS45T55 sample measured at 8.7\,K in the superconducting state. To fully resolve the Fermi surface and band structures, the ARPES measurements were carried out under three different laser light polarization geometries (See the laser \textit{s}, \textit{p} and circular polarizations in Fig. 1b). The momentum space covered by the 6.994\,eV laser in the entire Brillouin zone is shown in Fig. 1c. The  momentum space is large enough to cover the whole Fermi pockets around the $\Gamma$ point simultaneously under the same condition with very dense momentum points. The Fermi surface mappings measured under three different light polarizations are shown in Fig. 1d-f. It is clear that they exhibit rather different spectral weight distributions. In the \textit{s} polarization geometry in Fig. 1d, the measured Fermi surface mapping shows two strong spots located along the $\Gamma$-X direction. In the \textit{p} polarization geometry in Fig. 1e, the two strong spots change their orientation to be along the $\Gamma$-Y direction. Additional spectral weight emerges at the $\Gamma$ point. In the circular polarization geometry, the spectral weight distribution of the Fermi surface mapping can be considered as the superposition of the two mappings shown in Fig. 1d and e. These results indicate that strong photoemission matrix element effects are involved in the ARPES measurements of the Fe(Se,Te) sample. This provides a good opportunity to resolve the individual band structures by taking advantage of the matrix element effects.

In order to determine the detailed Fermi surface and electronic structure, Fig. 2 shows momentum dependence of the band structure for the FeS45T55 sample measured at 8.7 K in the superconducting state under three laser polarizations. The momentum cuts crossing the zone center with different Fermi surface angles are defined as $\theta$ in Fig. 2a-c. In the \textit{s} polarization geometry, one band is observed clearly for the momentum cut along the horizontal direction($\theta$=0$^\circ$) (Fig. 2d1). With the increase of the Fermi surface angle $\theta$, the spectral weight of this band decreases gradually and becomes nearly invisible for the momentum cuts with the Fermi surface angle above 60$^\circ$ (Fig. 2d5 to d7). At $\Gamma$ point, there is  no clear spectral weight observed for all the momentum cuts. In the \textit{p} polarization geometry, the momentum dependent evolution of the band structure with the Fermi surface angle is nearly the opposite. The band clearly observed in Fig. 2d1 becomes invisible along the horizontal direction ($\theta$=0$^\circ$, Fig. 2e1), but gets stronger gradually when the Fermi surface angle increases (Fig. 2e1 to e7). It is noted that a strong spot is observed just at the $\Gamma$ point near the Fermi level which is present for all the momentum cuts (Fig. 2e1 to e7). In the momentum cut along the $\Gamma$-Y direction in Fig. 2e7, a flat band-like feature emerges around the Fermi level. This flat band-like feature is similar to that observed in the previous reports\cite{lubashevsky2012shallow,rinott2017tuning}. In the case of the circular polarization, the flat band-like feature is always observed for all the momentum cuts (Fig. 2f1 to f7).

Considering the light polarizations in Fig. 1b and combining with the momentum dependence of the band structures with the Fermi surface angle, we conclude that there are two Fermi surface sheets around the zone center as marked by the dashed lines in Fig. 1d-f. The outer Fermi surface mainly consists of the d$_{xz}$/d$_{yz}$ orbital with the d$_{yz}$ orbital dominating along the $\Gamma$-X direction and the d$_{xz}$ orbital dominating along the $\Gamma$-Y direction\cite{zhang2018observation,zhang2019multiple}. The inner Fermi surface has d$_{z^2}$ or d$_{xz}$ orbital component because it is always visible under the p polarization geometry which has an out-of-plane component of the electric field vector as shown in Fig. 1b. This tiny Fermi surface originates from the topological surface state\cite{zhang2018observation,zhang2019multiple}.


Figure 3 shows the temperature dependence of the Fermi surface mappings and the band structures measured along $\Gamma$-X/Y directions in order to understand the origin of the flat band-like feature. Three typical momentum cuts are chosen to study the temperature dependence. For the Cut1 in the case of \textit{s} polarization (red line in Fig. 3d), only the outer d$_{yz}$ band is observed (Fig. 3g1-g6). For the Cut2 in the case of \textit{p} polarization (horizontal red line in Fig. 3e), the d$_{yz}$ band is absent and only the topological surface state near the Fermi level and the d$_{xz}$ band below the Fermi level are observed (Fig. 3h1-h6). For the Cut3 (vertical red line in Fig. 3e), all the d$_{yz}$ band, d$_{xz}$ band and the topological surface state are observed simultaneously (Fig. 3i1-i6). In the normal state (30\,K, 22\,K, 17\,K and 14.5\,K), both the d$_{yz}$ band and the topological surface state show relatively weak spectral weight around the Fermi level and there is no flat band-like feature observed around the Fermi level. With decreasing temperature into the superconducting state, the spectral weight around the Fermi level from these two bands starts to increase due to the formation of the Bogoliubov quasiparticles (12\,K and 8.7\,K as shown in Fig. 3g and h). For the Cut3 in the superconducting state, the Bogoliubov quasiparticles from these two bands combine to form a flat band-like feature as shown in Fig. 3i6.

To quantitatively describe the origin of the flat band-like feature, we replot the band structures both in the normal state and in the superconducting state along the momentum cuts from Cut1 to Cut3 (as marked in Fig. 3d and 3e) in Fig. 4 to Fig. 6, respectively. The corresponding EDC (energy distribution curve) stacks and the second derivative images with respect to the energy and momentum are also provided correspondingly, in order to extract the band dispersions quantitatively. Fig. 4a and 4b show band structures measured along Cut1 (Fig. 3d) in \textit{s} polarization geometry in the normal state (Fig. 4a) and the superconducting state (Fig. 4b). The band structures are divided by the corresponding Fermi–Dirac distribution functions to show the electronic states above the Fermi level. The observed bands are marked by the guided lines and mainly the d$_{yz}$ band is observed. Two bands are observed in the normal state, as marked by the blue and green lines (Fig. 4a). The upper one crosses the Fermi level while the lower one is below the Fermi level. In fact, these two bands have the same orbital component from the same original d$_{yz}$ band which splits into two bands because of its hybridization with the d$_{xy}$ band\cite{zhang2019multiple}. In the superconducting state, the upper band opens a superconducting gap and forms a regular Bogoliubov band as shown by the blue line in Fig. 4b. Fig. 4e plots quantitatively the dispersions of the upper d$_{yz}$ band in the normal state (red line) and in the superconducting state (blue circles). The dispersion in the superconducting state $\varepsilon_S$ can be well described from the normal state dispersion $\varepsilon_N$ by the BCS relation $\varepsilon_S=-\sqrt{\varepsilon_N^{2}+\Delta^{2}}$ with the superconducting gap $\Delta$=1.4\,meV. Therefore, the flat band-like feature in the superconducting state is not formed from the d$_{yz}$ band alone and the band structure in the superconducting state can be well described by the BCS picture.

Fig. 5a and 5b show band structures measured along Cut2 (Fig. 3e) in \textit{p} polarization geometry in the normal state (Fig. 5a) and the superconducting state (Fig. 5b). Combining the original images and the second derivative images in Fig. 5c-5d and Fig. 5e-5f, only the topological surface state (SS) is observed near the Fermi level (red line in Fig. 5a and 5b)\cite{zhang2018observation,zhang2019multiple}. In the normal state, the topological surface state is very close to the Fermi level (Fig. 5c). In the superconducting state, a section of flat band is formed around $\Gamma$ (Fig. 5d). Fig. 5i plots quantitatively the dispersions of the topological surface state in the normal state (red line) and in the superconducting state (blue circles) obtained from the EDC stacks in Fig. 5g and 5h. The dispersion in the superconducting state $\varepsilon_S$ can also be well described from the normal state dispersion $\varepsilon_N$ by the BCS relation $\varepsilon_S=-\sqrt{\varepsilon_N^{2}+\Delta^{2}}$ with the superconducting gap $\Delta$=1.4\,meV. Therefore, the flat band feature around $\Gamma$ in the superconducting state originates from the topological surface state. This is different from the previous report that the flat band around $\Gamma$ in the superconducting state is from the electron-like band above the Fermi level in the normal state\cite{okazaki2014superconductivity}.

Fig. 6a and 6b show band structures measured along Cut3 (Fig. 3e) in \textit{p} polarization geometry in the normal state (Fig. 6a) and the superconducting state (Fig. 6b). Combining the original images and the second derivative images in Fig. 6c-6d and Fig. 6e-6f, both the d$_{yz}$ band (blue and green lines in Fig. 6a and 6b) and the topological surface state (red line in Fig. 6a and 6b) are observed near the Fermi level. A flat band-like feature can be observed near the Fermi level in Fig. 6b which appears to be more obvious in the second derivative image in Fig. 6d. As analyzed in Fig. 4 and Fig. 5, it is now clear that this flat band-like feature is composed of two components: the d$_{yz}$ Bogoliubov bands on both sides and the topological surface state Bogoliubov band at the $\Gamma$ point (Fig. 6i). Therefore, it is not a complete flat band but consists of three discrete sections. Similar three sections are also present in the previous ARPES measurements\cite{okazaki2014superconductivity,rinott2017tuning}. These observations are clearly not consistent with the band structure expected from the BCS-BEC crossover picture (Fig 6j).

Based on the above results, we can now establish a coherent picture to understand the observed electronic structure of FeSe$_{0.45}$Te$_{0.55}$ superconductor around the zone center. There are two Fermi surface sheets around the $\Gamma$ point. The outer one is from bulk band which consists of d$_{yz}$ orbital along the $\Gamma$-X direction and d$_{xz}$ orbital along the $\Gamma$-Y direction. The inner tiny one is from non-trivial topological surface state which has d$_{z^2}$ or d$_{xz}$ orbital component. Along the $\Gamma$-X direction, only the d$_{yz}$ band can be observed in the s polarization geometry (Fig. 4) while in the p polarization geometry only the topological surface state can be observed near the Fermi level (Fig. 5). Along the $\Gamma$-Y direction in the p polarization geometry, both bands can be observed simultaneously (Fig. 6). In the normal state, both bands cross the Fermi level with a quite small Fermi energy of 2-3\,meV. In the superconducting state, the two bands both form Bogoliubov quasiparticles with the superconducting gap opening which can be well described by the BCS picture. In the superconducting state, the two kinds of Bogoliubov bands combine to form a flat band-like feature near the Fermi level (Fig. 6).


Our present results clearly rule out the existences of the BCS-BEC crossover in FeSe$_{0.45}$Te$_{0.55}$ superconductor. First, the unique flat band-like feature was considered to originate from the d$_{yz}$ band\cite{lubashevsky2012shallow,rinott2017tuning}. By isolating the d$_{yz}$ band in the s polarization geometry, our results indicate that the d$_{yz}$ band alone can not produce the flat band-like feature in the superconducting state (Fig. 4) and the band structure in the superconducting state can be described in the BCS picture. Second, the central flat band around $\Gamma$ originates from the topological surface state in the superconducting state (Fig. 5). The dispersion of this surface state in the superconducting state can be obtained from its normal dispersion in the BCS picture. This rules out the possibility of the BCS-BEC crossover for this topological surface state\cite{okazaki2014superconductivity}. Third, we have shown that the flat band-like feature is actually composed of three discrete sections of bands coming from both the d$_{yz}$ band and the topological surface state (Fig. 6). It can be well understood in the BCS picture and is not consistent with the expect band structure (Fig. 6j) in the BCS-BEC crossover picture.
\\

\noindent {\bf{4. Summary}}\\

In summary, we carried out high-resolution laser ARPES measurements on FeSe$_{0.45}$Te$_{0.55}$ superconductor. By taking advantage of the photoemission matrix element effects, the detailed electronic structure in the normal state and in the superconducting state are determined. We find that the d$_{yz}$ band alone does not form a flat band-like feature in the superconducting state and the measured dispersion can be well described by the BCS picture. The flat band-like feature is formed from the combination of the d$_{yz}$ band and the topological surface state band in the superconducting state. These results reveal the origin of the flat band-like feature and rule out the presence of BCS-BEC crossover in Fe(Se,Te) superconductor.
\\

\vspace{3mm}

\noindent {\bf Acknowledgements}\\

This work is supported by the National Key Research and Development Program of China (Grant Nos. 2021YFA1401800, 2022YFA1604200, 2022YFA1403900 and 2023YFA1406000), the National Natural Science Foundation of China (Grant Nos. 12488201, 12374066, 12074411 and 12374154), the Strategic Priority Research Program (B) of the Chinese Academy of Sciences (Grant No. XDB25000000 and XDB33000000), the Innovation Program for Quantum Science and Technology (Grant No. 2021ZD0301800), the Youth Innovation Promotion Association of CAS (Grant No. Y2021006) and the Synergetic Extreme Condition User Facility (SECUF).\\

\vspace{3mm}

\noindent {\bf Author Contributions}\\

X.J.Z., L.Z. and J.J.J. proposed and designed the research. Y.D.G. and Z.A.R. contributed in sample growth. C.H.Y., Y.J.S., Y.W.C., J.M.S., X.Z., H.C., T.M.M., X.L.R., B.L., W.P.Z., N.C., F.F.Z., S.J.Z., F.Y., Z.M.W., Q.J.P., Z.Y.X., H.Q.M., G.D.L., L.Z. and X.J.Z. contributed to the development and maintenance of the ARPES system and related software development. J.J.J. carried out the ARPES experiment. J.J.J., L.Z. and X.J.Z. analyzed the data. J.J.J., L.Z. and X.J.Z. wrote the paper. All authors participated in discussion and comment on the paper.
\\

\vspace{3mm}

{\bf{References}}\\


%
%
%

\newpage

\begin{figure*}[tbp]
\begin{center}
\includegraphics[width=1.0\columnwidth,angle=0]{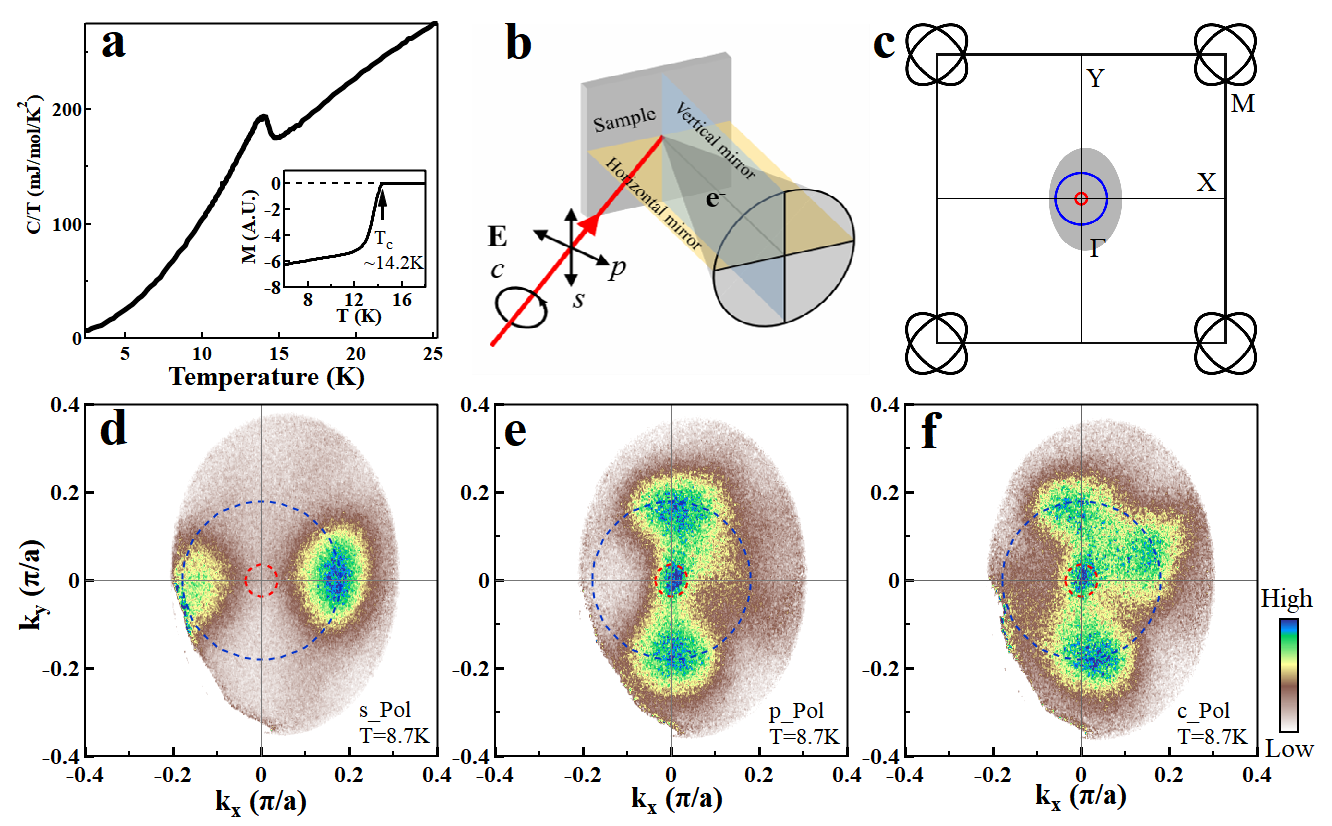}
\end{center}

\caption{{\bf Fermi surface of FeSe$_{0.45}$Te$_{0.55}$ measured under different polarization geometries.} (a) Temperature dependence of specific heat plotted as C/T vs T for the samples with 0.02 Mn doping. The inset shows magnetic measurement of the sample. The onset T$_c$  is 14.2\,K with a transition width of $\sim$1.5\,K\cite{ZARen2022YDGu}. (b) Schematic illustration of the polarization geometries used in the ARPES measurements. The laser light is incident on the sample 50 degrees with respect to the normal direction of the sample surface. The electric field vector of the s-polarization is vertical while it is horizontal for the p polarization. In the s polarization, the electric field is fully in the sample plane while in the p polarization, it has both in-plane and out-of-plane components. The c polarization refers to the circular polarization geometry. (c) Schematic Fermi surface of Fe(Se,Te). The central shaded area represents the momentum region in our ARPES measurements. (d-f) Fermi surface mapping of FeSe$_{0.45}$Te$_{0.55}$ measured at 8.7 K under the s polarization (d), p polarization (e) and circular polarization (f). The blue and red dashed lines represent the measured Fermi surface around $\Gamma$ point.
}
\end{figure*}

\begin{figure*}[tbp]
\begin{center}
\includegraphics[width=1.0\columnwidth,angle=0]{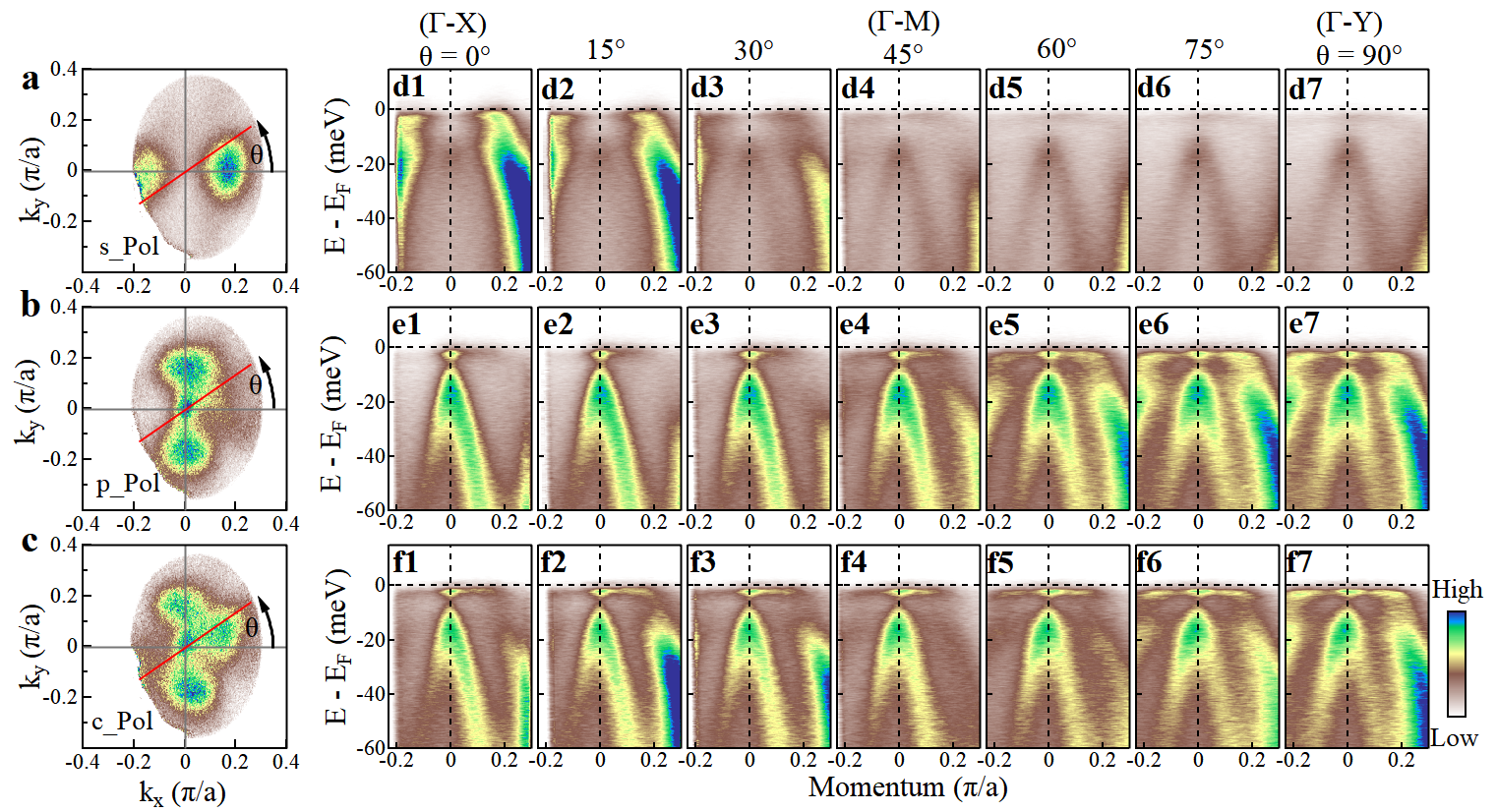}
\end{center}

\caption{{\bf Band structures of FeSe$_{0.45}$Te$_{0.55}$ measured at 8.7\,K along different momentum cuts under three different polarization geometries. } (a-c) Fermi surface mappings of FeSe$_{0.45}$Te$_{0.55}$ measured under s, p and c polarization geometries, respectively. (d1-d7, e1-e7, f1-f7) Measured band structures along different momentum cuts. The momentum cuts are represented by the angle $\theta$ which is defined by the red lines in a-c. The momentum cuts of $\theta$ =0, 45 and 90 correspond to the direction of $\Gamma$-X, $\Gamma$-M and $\Gamma$-Y, respectively.
}
\end{figure*}

\begin{figure*}[tbp]
\begin{center}
\includegraphics[width=1.0\columnwidth,angle=0]{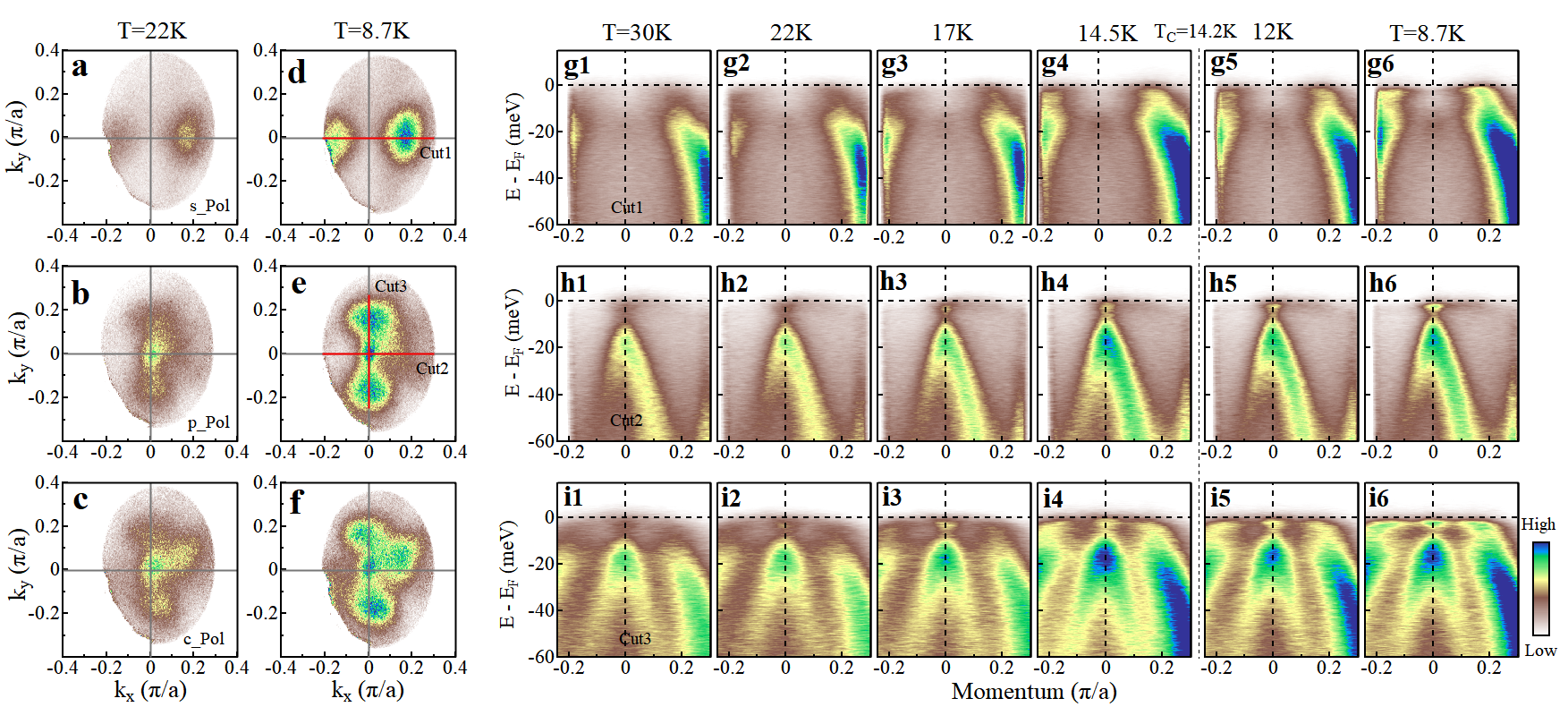}
\end{center}

\caption{{\bf Fermi surface mappings and band structures of FeSe$_{0.45}$Te$_{0.55}$ measured at different temperatures under three different polarization geometries}  (a-c) Fermi surface of FeSe$_{0.45}$Te$_{0.55}$ measured at 22\,K in the normal state under s, p and c polarization geometries. (d-f) Same as (a-c) but measured at 8.7 K in the superconducting state. (g1-g6, h1-h6, i1-i6) Band structures measured at different temperatures along momentum Cut1 (g), Cut2 (h) and Cut3 (i). The location of Cut1 is shown by the red line in (d), while the location of Cut2 and Cut3 is shown by the red lines in (e).
}
\end{figure*}

\begin{figure*}[tbp]
\begin{center}
\includegraphics[width=1.0\columnwidth,angle=0]{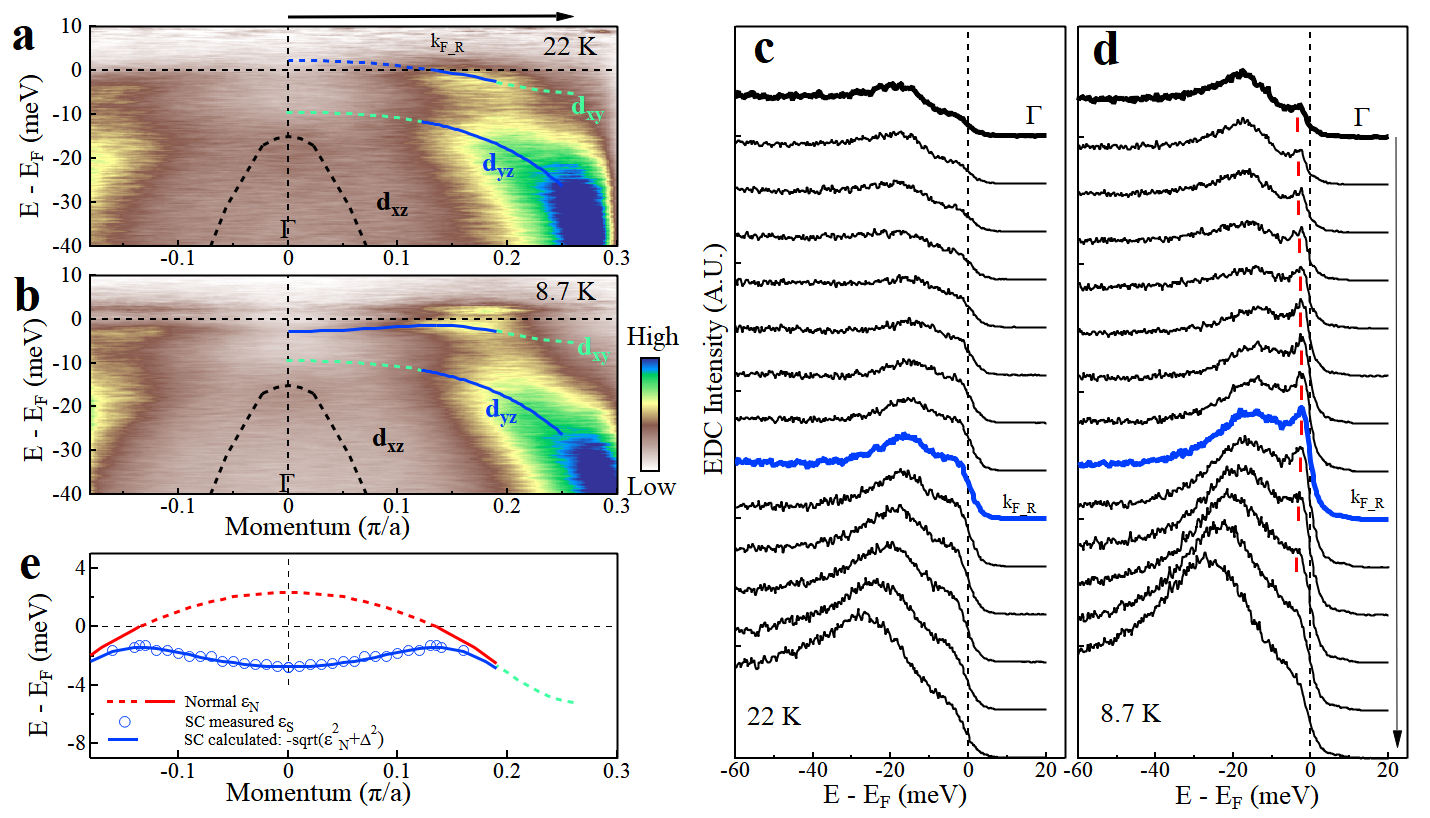}
\end{center}

\caption{{\bf Detailed analysis of the d$_{yz}$ band measured in the normal state and the superconducting state. } (a-b) Band structures measured at 22\,K in the normal state (a) and at 8.7\,K in the superconducting state (b). These data are taken around $\Gamma$ point along $\Gamma$-X direction under the s polarization geometry. The images are divided by the corresponding Fermi–Dirac distribution functions in order to show the electronic states above the Fermi level. The observed d$_{xz}$, d$_{yz}$ and d$_{xy}$ bands are marked and labeled. (c-d) EDC stacks in the normal state and in the superconducting state obtained from the original band structures in (a) and (b), respectively, without removing the Fermi–Dirac distribution functions. The corresponding momentum range is marked by the arrowed line on top of (a). The red tips in (d) mark the EDC peak positions. (e) The band dispersion in the normal state ($\epsilon_N$) obtained from (a) (red solid line) and the band dispersion in the superconducting state ($\epsilon_S$) obtain from (b) and (d) (blue circles). The normal state band dispersion above the Fermi level (red dashed line) is extrapolated by assuming a parabolic relation. The blue line is calculated from the normal state band dispersion by the BCS relation -$\sqrt{\epsilon_N^2+\Delta^2}$ with the superconducting gap $\Delta$=1.4\,meV.
}

\end{figure*}

\begin{figure*}[tbp]
\begin{center}
\includegraphics[width=1.0\columnwidth,angle=0]{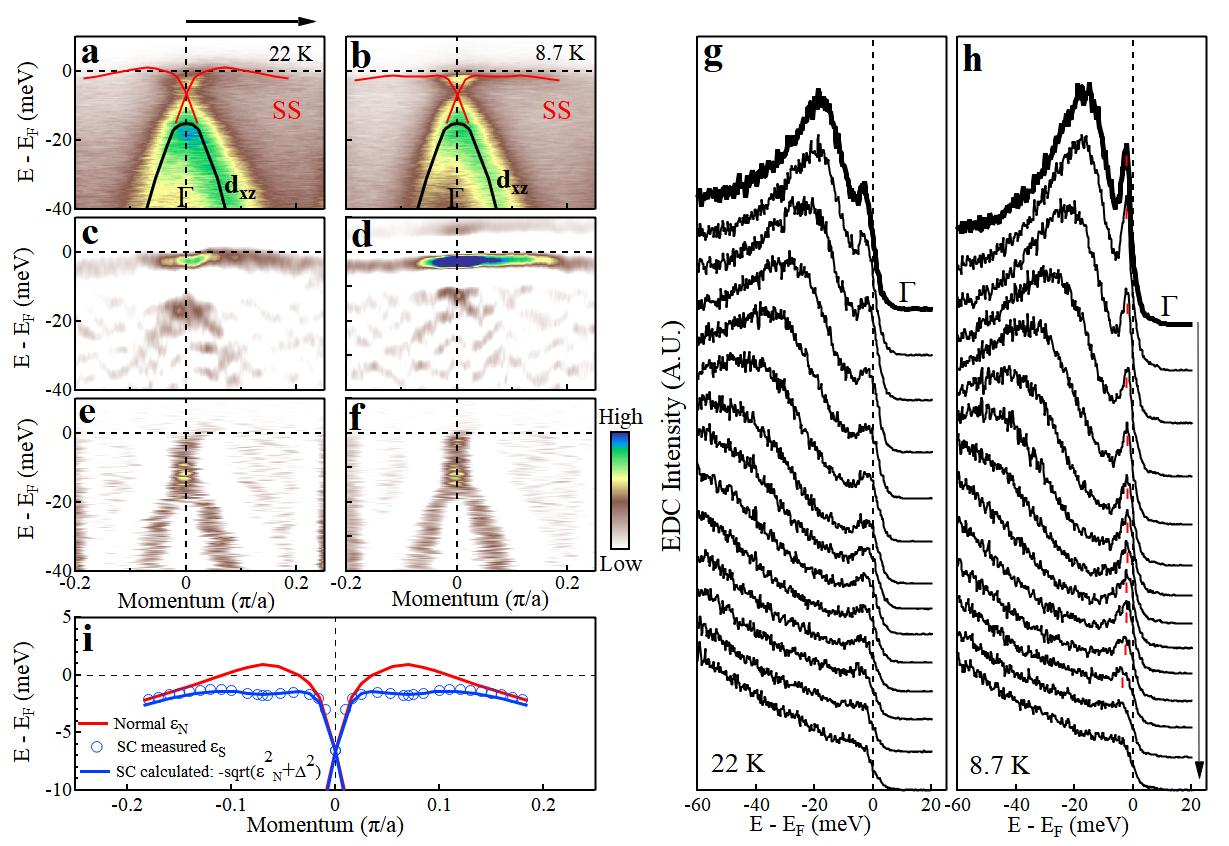}
\end{center}

\caption{{\bf Detailed analysis of the topological surface state band measured in the normal state and the superconducting state. } (a-b) Band structures measured at 22\,K in the normal state (a) and at 8.7\,K in the superconducting state (b). These data are taken around $\Gamma$ point along $\Gamma$-X direction under the p polarization geometry. The observed d$_{xz}$ and topological surface state (SS) bands are marked and labeled. (c-d) Corresponding EDC second derivative images obtained from (a-b). (e-f) Corresponding MDC second derivative images obtained from (a-b). (g-h) EDC stacks in the normal state and in the superconducting state obtained from (a) and (b), respectively. The corresponding momentum range is marked by the arrowed line on top of (a). The red tips in (h) mark the EDC peak positions. (i) The band dispersion of the topological surface state in the normal state ($\epsilon_N$) obtained from (a) (red solid line) and the band dispersion in the superconducting state ($\epsilon_S$) obtain from (b) and (h) (blue circles). The blue line is calculated from the normal state band dispersion by the BCS relation -$\sqrt{\epsilon_N^2+\Delta^2}$ with the superconducting gap $\Delta$=1.4\,meV.
}
\end{figure*}

\begin{figure*}[tbp]
\begin{center}
\includegraphics[width=1.0\columnwidth,angle=0]{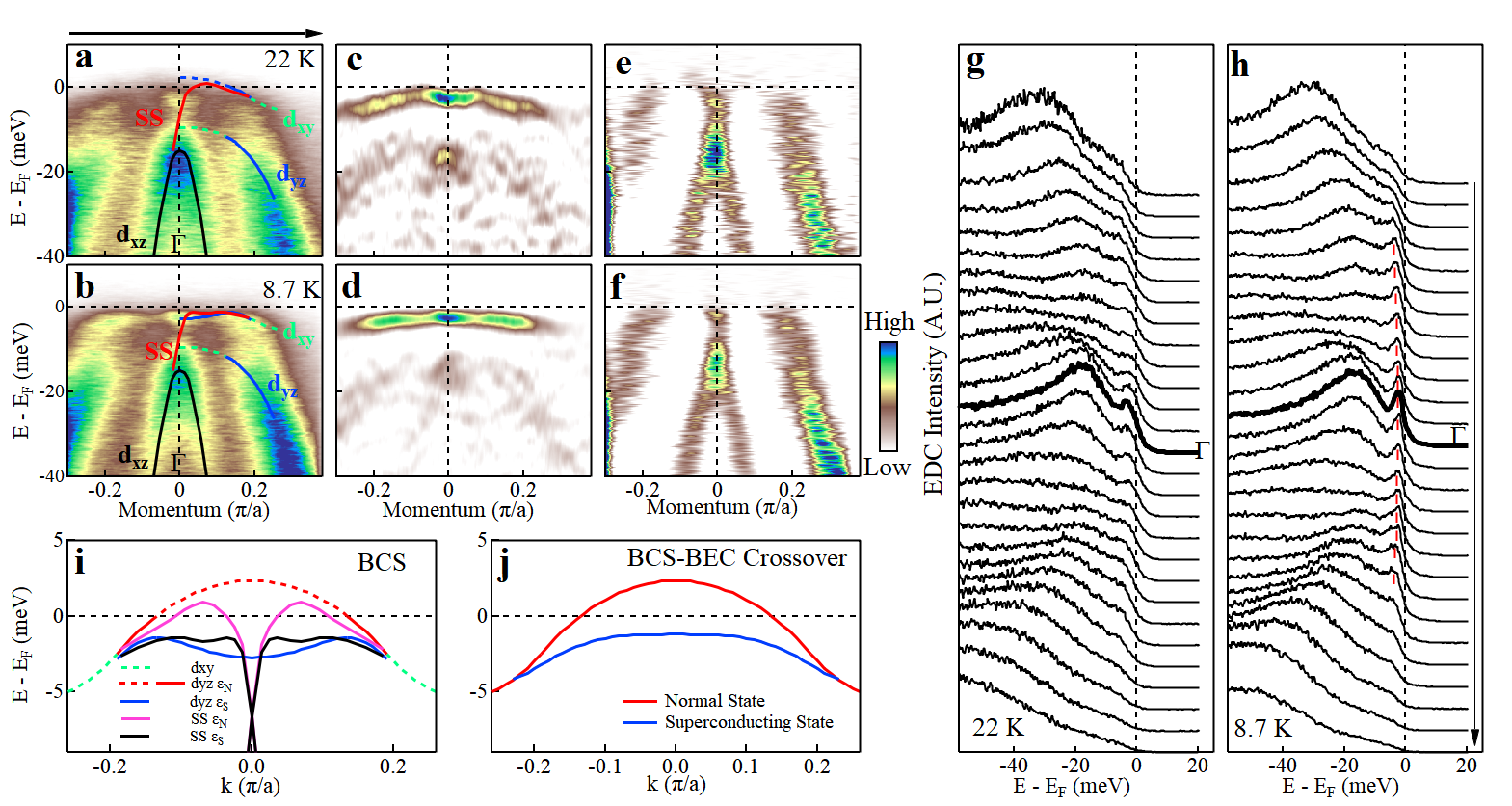}
\end{center}

\caption{{\bf Detailed analysis of the origin of the flat band-like feature observed in the superconducting state. } (a-b) Band structures measured at 22\,K in the normal state (a) and at 8.7\,K in the superconducting state (b). These data are taken around $\Gamma$ point along $\Gamma$-Y direction under the p polarization geometry. In this case, all the d$_{xz}$, d$_{yz}$, d$_{xy}$ and topological surface state bands can be observed which are marked and labeled. (c-d) Corresponding EDC second derivative images obtained from (a-b). (e-f) Corresponding MDC second derivative images obtained from (a-b). (g-h) EDC stacks in the normal state and in the superconducting state obtained from (a) and (b), respectively. The corresponding momentum range is marked by the arrowed line on top of (a). The red tips in (h) mark the EDC peak positions. (i) The band dispersions of the d$_{yz}$ band in the normal state (red line) and in the superconducting state (blue line) obtained from Fig. 4e. The band dispersions of the topological surface state band in the normal state (pink line) and in the superconducting state (black line) obtained from Fig. 5i are also plotted. (j) Schematic illustration of the band structures in the normal and superconducting states in the BCS-BEC crossover picture.
}

\end{figure*}


\begin{thebibliography}{10}

\bibitem{JRSchrieffer1957JBardeen}
J.~Bardeen, L.~N. Cooper, and J.~R. Schrieffer.
\newblock {Theory of Superconductivity}.
\newblock {\em Physical Review}, 108(5):1175, 1957.

\bibitem{SStringari1999FDalfovo}
Franco Dalfovo, Stefano Giorgini, Lev~P. Pitaevskii, and Sandro Stringari.
\newblock {Theory of Bose-Einstein condensation in trapped gases}.
\newblock {\em Reviews of Modern Physics}, 71(3):463--512, 1999.

\bibitem{SStringari2008SGiorgini}
Stefano Giorgini, Lev~P. Pitaevskii, and Sandro Stringari.
\newblock {Theory of ultracold atomic Fermi gases}.
\newblock {\em Reviews of Modern Physics}, 80(4):1215--1274, 2008.

\bibitem{eagles1969possible}
DM~Eagles.
\newblock Possible pairing without superconductivity at low carrier
  concentrations in bulk and thin-film superconducting semiconductors.
\newblock {\em Physical Review}, 186(2):456, 1969.

\bibitem{leggett2008diatomic}
Anthony~James Leggett.
\newblock Diatomic molecules and cooper pairs.
\newblock In {\em Modern Trends in the Theory of Condensed Matter: Proceedings
  of the XVI Karpacz Winter School of Theoretical Physics, February 19--March
  3, 1979 Karpacz, Poland}, pages 13--27. Springer, 2008.

\bibitem{nozieres1985bose}
Ph~Nozieres and S~Schmitt-Rink.
\newblock Bose condensation in an attractive fermion gas: From weak to strong
  coupling superconductivity.
\newblock {\em Journal of Low Temperature Physics}, 59:195--211, 1985.

\bibitem{KLevin2005QJChen}
Qijin Chen, Jelena Stajic, Shina Tan, and K.~Levin.
\newblock {BCS–BEC crossover: From high temperature superconductors to
  ultracold superfluids}.
\newblock {\em Physics Reports}, 412(1):1--88, 2005.

\bibitem{randeria2014crossover}
Mohit Randeria and Edward Taylor.
\newblock Crossover from bardeen-cooper-schrieffer to bose-einstein
  condensation and the unitary fermi gas.
\newblock {\em Annu. Rev. Condens. Matter Phys.}, 5(1):209--232, 2014.

\bibitem{chen2024superconductivity}
Qijin Chen, Zhiqiang Wang, Rufus Boyack, Shuolong Yang, and K~Levin.
\newblock When superconductivity crosses over: From bcs to bec.
\newblock {\em Reviews of Modern Physics}, 96(2):025002, 2024.

\bibitem{lubashevsky2012shallow}
Y~Lubashevsky, E~Lahoud, K~Chashka, D~Podolsky, and A~Kanigel.
\newblock Shallow pockets and very strong coupling superconductivity in fese x
  te1- x.
\newblock {\em Nature Physics}, 8(4):309--312, 2012.

\bibitem{kasahara2014field}
Shigeru Kasahara, Tatsuya Watashige, Tetsuo Hanaguri, Yuhki Kohsaka, Takuya
  Yamashita, Yusuke Shimoyama, Yuta Mizukami, Ryota Endo, Hiroaki Ikeda,
  Kazushi Aoyama, et~al.
\newblock Field-induced superconducting phase of fese in the bcs-bec
  cross-over.
\newblock {\em Proceedings of the National Academy of Sciences},
  111(46):16309--16313, 2014.

\bibitem{okazaki2014superconductivity}
Kozo Okazaki, Yoshiaki Ito, Yuichi Ota, Yoshinori Kotani, Takahiro Shimojima,
  Takayuki Kiss, Shuntaro Watanabe, C-T Chen, Seiji Niitaka, Tetsuo Hanaguri,
  et~al.
\newblock Superconductivity in an electron band just above the fermi level:
  possible route to bcs-bec superconductivity.
\newblock {\em Scientific reports}, 4(1):4109, 2014.

\bibitem{rinott2017tuning}
Shahar Rinott, KB~Chashka, Amit Ribak, Emile~DL Rienks, Amina Taleb-Ibrahimi,
  Patrick Le~Fevre, Fran{\c{c}}ois Bertran, Mohit Randeria, and Amit Kanigel.
\newblock Tuning across the bcs-bec crossover in the multiband superconductor
  ${\text{fe}}_{1+y}{\text{se}}_x{\text{te}}_{1-x}$: An angle-resolved
  photoemission study.
\newblock {\em Science advances}, 3(4):e1602372, 2017.

\bibitem{hashimoto2020bose}
Takahiro Hashimoto, Yuichi Ota, Akihiro Tsuzuki, Tsubaki Nagashima, Akiko
  Fukushima, Shigeru Kasahara, Yuji Matsuda, Kohei Matsuura, Yuta Mizukami,
  Takasada Shibauchi, et~al.
\newblock Bose-einstein condensation superconductivity induced by disappearance
  of the nematic state.
\newblock {\em Science Advances}, 6(45):eabb9052, 2020.

\bibitem{ZARen2022YDGu}
Yadong Guu, Menghu Zhou, Mengdi Zhang, Yanwei Wu, Binbin Ruan, Xingyuan Hou,
  Fan Zhang, Peijie Jiang, Qingsong Yang, Geng Li, Mingwei Ma, Genfu Chen, Lei
  Shan, and Zhian Ren.
\newblock {Bulk superconductivity in one-step grown Fe(Te,Se) crystals free of
  interstitial iron by minor Mn doping}.
\newblock {\em Science China Materials}, 65(9):2472--2478, 2022.

\bibitem{XJZhou2018}
Xingjiang Zhou, Shaolong He, Guodong Liu, Lin Zhao, Li~Yu, and Wentao Zhang.
\newblock {New developments in laser-based photoemission spectroscopy and its
  scientific applications: a key issues review}.
\newblock {\em Reports on Progress in Physics}, 81(6):062101, 2018.

\bibitem{zhang2018observation}
Peng Zhang, Koichiro Yaji, Takahiro Hashimoto, Yuichi Ota, Takeshi Kondo, Kozo
  Okazaki, Zhijun Wang, Jinsheng Wen, Genda~D Gu, Hong Ding, et~al.
\newblock Observation of topological superconductivity on the surface of an
  iron-based superconductor.
\newblock {\em Science}, 360(6385):182--186, 2018.

\bibitem{zhang2019multiple}
Peng Zhang, Zhijun Wang, Xianxin Wu, Koichiro Yaji, Yukiaki Ishida, Yoshimitsu
  Kohama, Guangyang Dai, Yue Sun, Cedric Bareille, Kenta Kuroda, et~al.
\newblock Multiple topological states in iron-based superconductors.
\newblock {\em Nature Physics}, 15(1):41--47, 2019.

\end{thebibliography}
\end{document}